\documentclass[graybox,nospthms]{svmult}

\usepackage{mathptmx}       
\usepackage{helvet}         
\usepackage[scaled]{inconsolata}        
\usepackage{type1cm}        
\usepackage{makeidx}         
\usepackage{graphicx}        
\usepackage{multicol}        
\usepackage[bottom]{footmisc}

\usepackage{mathpartir}

\usepackage{alltt}      
\usepackage{amsfonts}   
\usepackage{amsmath}    

\usepackage{amssymb}

\usepackage{amsthm}     
\usepackage{bussproofs}
\EnableBpAbbreviations
\usepackage{cite}
\usepackage{combelow}
\usepackage{cprotect}
\usepackage{datetime}
\usepackage[inline]{enumitem}
\usepackage[colorlinks,linkcolor=black,citecolor=black,urlcolor=black]{hyperref}
\usepackage[T1]{fontenc}
\usepackage[utf8]{inputenc}
\usepackage{isabelle}
\usepackage{isabellesym}
\usepackage{listings}
\usepackage{makeidx}
\usepackage{mathpartir}
\usepackage{multirow}
\usepackage{newunicodechar}
\usepackage{proof}
\usepackage{stmaryrd}
\usepackage{tikz}
\usepackage{tikz-cd}
\usepackage{tikz-qtree}
\usepackage{upgreek}
\usepackage{varwidth}
\usepackage{xspace}
\usepackage[all]{xy}
\usepackage[all]{xypic}

\makeindex

\newunicodechar{∷}{::}
\newunicodechar{∀}{\ensuremath{\forall}}
\newunicodechar{∃}{\ensuremath{\exists}}
\newunicodechar{Σ}{\ensuremath{\Sigma}}
\newunicodechar{Γ}{\ensuremath{\Gamma}}
\newunicodechar{Δ}{\ensuremath{\Delta}}
\newunicodechar{∨}{\ensuremath{\lor}}
\newunicodechar{∧}{\ensuremath{\land}}
\newunicodechar{¬}{\ensuremath{\neg}}
\newunicodechar{λ}{\ensuremath{\lambda}}
\newunicodechar{∈}{\ensuremath{\in}}
\newunicodechar{∉}{\ensuremath{\not\in}}
\newunicodechar{∗}{\ensuremath{\ast}}
\newunicodechar{ℕ}{\ensuremath{\mathbb{N}}}
\newunicodechar{ℤ}{\ensuremath{\mathbb{Z}}}
\newunicodechar{ℚ}{\ensuremath{\mathbb{Q}}}
\newunicodechar{ℝ}{\ensuremath{\mathbb{R}}}
\newunicodechar{ℂ}{\ensuremath{\mathbb{C}}}
\newunicodechar{≤}{\ensuremath{\le}}
\newunicodechar{≺}{\ensuremath{\prec}}
\newunicodechar{∧}{\ensuremath{\wedge}}
\newunicodechar{⇒}{\ensuremath{\Rightarrow}}
\newunicodechar{→}{\ensuremath{\to}}
\newunicodechar{≠}{\ensuremath{\neq}}
\newunicodechar{⋯}{\ensuremath{\cdots}}
\newunicodechar{α}{\ensuremath{\alpha}}
\newunicodechar{β}{\ensuremath{\beta}}
\newunicodechar{γ}{\ensuremath{\gamma}}
\newunicodechar{δ}{\ensuremath{\delta}}
\newunicodechar{ε}{\ensuremath{\varepsilon}}
\newunicodechar{φ}{\ensuremath{\phi}}
\newunicodechar{×}{\ensuremath{\times}}
\newunicodechar{₀}{\ensuremath{_{0}}}
\newunicodechar{₁}{\ensuremath{_{1}}}
\newunicodechar{₂}{\ensuremath{_{2}}}
\newunicodechar{₃}{\ensuremath{_{3}}}
\newunicodechar{₄}{\ensuremath{_{4}}}
\newunicodechar{₅}{\ensuremath{_{5}}}
\newunicodechar{₆}{\ensuremath{_{6}}}
\newunicodechar{₇}{\ensuremath{_{7}}}
\newunicodechar{₈}{\ensuremath{_{8}}}
\newunicodechar{₉}{\ensuremath{_{9}}}
\newunicodechar{⟨}{\ensuremath{\langle}}
\newunicodechar{⟩}{\ensuremath{\rangle}}
\newunicodechar{⟦}{\ensuremath{\llbracket}}
\newunicodechar{⟧}{\ensuremath{\rrbracket}}

\DeclareSymbolFont{letters}{OML}{txmi}{m}{it}

\usetikzlibrary{petri} 
\usetikzlibrary{er}    
\usetikzlibrary{positioning,calc}

\makeatletter
\providecommand*{\toclevel@titlech}{0}
\edef\toclevel@authorch{\the\numexpr\toclevel@titlech+3}
\makeatother

\theoremstyle{plain}
\newtheorem{theorem}{Theorem}[chapter]

\theoremstyle{definition}
\newtheorem{definition}[theorem]{Definition}
\newtheorem{example}[theorem]{Example}

\newcommand\authorsandsecondreaders[2]{\authorrunning{#1}\tocauthor{#1}\author{#1 \\~\\~\\ Second readers: #2}}

\newcommand\Forall[2]{\forall #1.\; #2}

\newcommand\Or{\mathrel\lor}

\newcommand{\fn}[1]{\ensuremath{\mathsf{#1}}}     

\DeclareSymbolFont{letters}{OML}{txmi}{m}{it} 
\let\oldPhi=\Phi
\renewcommand\Phi{\mathrm{\oldPhi}}

\DeclareUnicodeCharacter{3C3}{\ensuremath{\sigma}}
\DeclareUnicodeCharacter{220F}{\ensuremath{\Pi}}
\DeclareUnicodeCharacter{03A0}{\ensuremath{\Pi}}
\DeclareUnicodeCharacter{3B5}{\ensuremath{\varepsilon}}
\DeclareUnicodeCharacter{2211}{\ensuremath{\Sigma}}
\DeclareUnicodeCharacter{2124}{\ensuremath{\mathbb{Z}}}
\DeclareUnicodeCharacter{2208}{\ensuremath{\in}}
\DeclareUnicodeCharacter{27F9}{\ensuremath{\Longrightarrow}}
\DeclareUnicodeCharacter{21D2}{\ensuremath{\Rightarrow}}
\DeclareUnicodeCharacter{03BB}{\ensuremath{\lambda}}

\begin{document}

\setcounter{chapter}{19}
\title{Verification of Provers and Solvers}
\authorsandsecondreaders{Ren\'e~Thiemann}{Fr\'ed\'eric Blanqui and Peter Lammich}
\authorrunning{Ren\'e~Thiemann}

\institute{%
Ren\'e Thiemann \at University of Innsbruck, Innsbruck, Austria, \email{rene.thiemann@uibk.ac.at}
}
%
%
\maketitle
\label{chap:verification-of-provers-and-solvers}

\newcommand\thmref[1]{Theorem~\ref{#1}}
\newcommand\exref[1]{Example~\ref{#1}}
\newcommand\defref[1]{Definition~\ref{#1}}

\newcommand\iirule[3]{\inferrule*[right={\normalfont{#3}}]{#1}{#2}}
\newcommand\trssym[1]{\fn{#1}}
\newcommand\Fack{\trssym{ack}}
\newcommand\Fs{\trssym{s}}
\newcommand\Ff{\trssym{f}}
\newcommand\Fnil{\trssym{nil}}
\newcommand\Fz{\trssym{0}}
\newcommand\RR{{\cal R}}

\index{Thiemann, Ren\'e}
\index{Blanqui, Fr\'ed\'eric}
\index{Lammich, Peter}

\abstract{
Automatic deduction tools such as automatic theorem provers, SAT
(satisfiability) solvers, SMT (satisfiability modulo theories) solvers, and
termination analyzers can be connected to proof assistants using various
approaches, notably by certification and verification. This chapter reviews and
compares the
approaches available, and mentions several successful applications. 
}
\bigskip
\bigskip

\section{Introduction}

In this chapter, we review the connection between proof assistants and
automatic deduction tools such as automatic theorem provers,
\index{SAT solving}
\index{SMT solving}
SAT (satisfiability) solvers, SMT (satisfiability modulo theories\index{satisfiability modulo theories}) solvers, and
\index{termination}
termination analyzers.
We will see how the reliability of automated tools
is increased with the help of proof assistants. Moreover, we
will see that proof assistants can be useful when extending
or modifying those inference rules that form the basis of the deduction tool
(e.g., the rules of CDCL, DPLL($T$), or the superposition calculus). There is
also the important
possibility of using deduction tools to improve automation in proof assistants,
which was already covered in Chapter~8. 

Although concrete domains such as SAT, SMT, and termination analysis are mentioned,
we do not assume any domain knowledge in these areas, but try to stay at a general
level. Still, we will use a running example in a concrete domain. To be more precise,
we consider a termination technique from first-order term rewriting that is also used in first-order theorem provers,
namely, the lexicographic path ordering\index{lexicographic path ordering|seealso{LPO}}\index{LPO}
(LPO) \cite{KaminLevy80}
and the recursive path ordering\index{recursive path ordering|seealso{RPO}}\index{RPO} with status (RPO) \cite{DBLP:journals/jsc/Dershowitz87}
where the latter ordering combines LPO and the recursive path ordering without status \cite{DBLP:journals/tcs/Dershowitz82}.

\index{LPO|(}
\begin{definition}
\label{LPO}
Let $\mathrm{\Sigma}$ be a signature of first-order terms, and $>$ be a strongly normalizing order on $\mathrm{\Sigma}$, called precedence.
Then the LPO induced by the precedence $>$ is defined as a binary relation $\succ$ on terms
via the following three inference rules:
\begin{gather*}
\label{sub}\iirule{s_i \succ t \Or s_i = t}{f(s_1,\dots,s_i,\dots,s_n) \succ t}{} \tag{sub} \\
\iirule{f > g \qquad \Forall{j \in \{1,\dots,m\}}{f(s_1,\dots,s_n) \succ t_j}}{f(s_1,\dots,s_n) \succ g(t_1,\dots,t_m)}{}
\tag{prec}\\
\label{lex}\iirule{s_i \succ t_i \qquad \Forall{j \in \{i+1,\dots,n\}}{f(s_1,\dots,s_n) \succ t_j}}%
{f(s_1,\dots,s_n) \succ f(s_1,\dots,s_{i-1},t_i,\dots,t_n)}{} \tag{lex}
\end{gather*}
\end{definition}

\index{RPO|(}

RPO extends LPO by a status where for every symbol one can choose whether the
arguments should be compared lexicographically as in (\ref{lex}) or via a
multiset comparison. We omit a precise definition of RPO and further term
rewriting related notions, since they are not important for understanding this
chapter. The interested reader is referred to a
textbook~\cite{DBLP:books/daglib/0092409} for details.

The following theorems are what make these orderings useful.

\begin{theorem}
\label{LPO sound}
LPO and RPO are reduction orders.
\end{theorem}

\begin{theorem}
\label{redorders}
Let $\succ$ be a reduction order.
Whenever $\ell \succ r$ for all rules $\ell \to r$ of a term rewrite system $\RR$,
then $\RR$ is terminating. 
\end{theorem}

\index{LPO|)}
\index{RPO|)}

\section{Certification}
\label{sect:certification}

\index{certification|(}

\emph{Certification} is an approach to validate results from unreliable sources,
such as highly optimized automated deduction tools. These tools may depend on other unverified
tools. Consider a termination prover that searches for suitable precedences of LPO with the help 
of an SMT solver for integer arithmetic\index{arithmetic},\footnote{A precedence can be synthesized by assigning natural numbers to each symbol,
and then $f > g$ is defined via a comparison of the numbers of $f$ and $g$.}
which itself relies on a C library for arbitrary precision integers.
We only assume that the tools generate some certificate in addition to the Boolean answer such as
``the input
is terminating.''\footnote{These kinds of tools are called certifying algorithms
or certifying computations by some authors~\cite{DBLP:journals/jar/AlkassarBMR14}. Here, we instead use
the notion of \relax{certification} exclusively
for the process or algorithm that checks a certificate, and not for the certificate-producing tools.}
Such a certificate usually consists of a proof sketch (which techniques or inference rules are applied in which order), 
in combination with auxiliary informations how the techniques are applied or parameterized.
Certificates can be quite verbose, but also sparse; for example, a certificate might only mention that
a certain decision procedure is applicable, without giving any further details.

The advantage of certification (checking the result) over verification (fully verifying the
automated tool)
is that the former is often simpler. For example, searching a suitable precedence
for an LPO or RPO termination proof is $\mathbf{NP}$-complete~\cite{DBLP:journals/tcs/KrishnamoorthyN85}, whereas checking an LPO or RPO proof with given
precedence is usually simple: polynomial time implementations of textbook definitions of LPO and RPO can easily be obtained by using 
memoization\index{memoization}.\footnote{Note that the complexity of checking RPO constraints for a given precedence depends on the exact definition of RPO. 
For certain extensions of RPO, already constraint checking is $\mathbf{NP}$-complete \cite{DBLP:conf/rta/ThiemannAN12}.}

\index{Ackermann--Péter function|(}

\begin{example}
\label{ex ackermann}
Consider the following encoding of the Ackermann--Péter function as a term rewrite system
over the signature $\Sigma = \{\Fack,\Fs,\Fz\}$.
\begin{align*}
\Fack(\Fz,m) &\to \Fs(m)\\
\Fack(\Fs(n),\Fz) &\to \Fack(n,\Fs(\Fz))\\
\Fack(\Fs(n),\Fs(m)) &\to \Fack(n,\Fack(\Fs(n),m))
\end{align*}
A basic certificate of termination is ``LPO with precedence $\Fack > \Fs$.''
An extended certificate might additionally contain inference trees for each rule, e.g.,
the tree in Figure~\ref{fig:inf-tree} for the second rule.
\begin{figure}[t]
\begin{equation*}
\iirule{\strut\iirule{\strut n = n}{\strut \Fs(n) \succ n}{(sub)}\\
\iirule{\strut\Fack > \Fs\\
\iirule{\strut\Fz = \Fz}{\strut\Fack(\Fs(n),\Fz) \succ \Fz}{(sub)}
}{\strut\Fack(\Fs(n),\Fz) \succ \Fs(\Fz)}{(prec)}
}{\strut\Fack(\Fs(n),\Fz) \succ \Fack(n,\Fs(\Fz))}{(lex)}
\end{equation*}
%
%
\caption{
\label{fig:inf-tree}
Inference tree for $\Fack(\Fs(n),\Fz) \succ \Fack(n,\Fs(\Fz))$}
\end{figure}
\end{example}

\index{Ackermann--Péter function|)}

For certifying proofs that are generated by deduction tools there are two essential questions.
\begin{enumerate}[label=(\Alph*)]
\item \label{thms} Are the applied techniques sound? \\
  For \exref{ex ackermann}, this corresponds to the question whether we believe that Theorems~\ref{LPO sound}~and~\ref{redorders} are indeed provable theorems.  
  For other kinds of proofs, we might want to know whether the inference rules of CDCL\index{CDCL} are sound, whether a given superposition\index{superposition} calculus is sound, etc.
\smallskip
\item \label{appl} Are the techniques applied correctly? \\
  In \exref{ex ackermann}, we have to ensure that the specified precedence is strongly normalizing and we need to check
  that $\ell \succ r$ is satisfied for each rule $\ell \to r$ of the rewrite system. Again, similar questions arise for other kinds of proofs. 
\end{enumerate} 

In the following, we will discuss four different ways to perform certification.

\subsection{Manual Certification}

Clearly, humans can answer both questions \ref{thms} and \ref{appl} for \exref{ex ackermann}: regarding \ref{thms}, the proofs can be found in the original
paper or in a textbook on rewriting, and these proofs can be checked manually; similarly, it is quite easy to verify manually that 
the inference tree in Figure~\ref{fig:inf-tree} is correct. We leave it as an exercise to the reader to figure out suitable inference trees 
for the other two rules. Strong normalization of the precedence of the example is obvious.

The problem with manual certification is that it is error-prone and inefficient and does not scale. In particular, answering question \ref{appl} 
without computer support is often feasible only for toy examples.

\subsection{Trusted Checker}
\label{subsect:trusted-checkers}

A better solution to question \ref{appl} is by automating this task.
The idea is the use of a trusted checker, a program which takes certificates such as the one in \exref{ex ackermann} as input, and then
validates whether all techniques within that proof have been applied correctly. 
Examples of such checkers have been used in various competitions---e.g., DRAT-trim\index{DRAT-trim}~\cite{DBLP:conf/sat/WetzlerHH14} was used as a checker in
the SAT competition, and CPAchecker\index{CPAchecker} \cite{DBLP:conf/cav/BeyerK11} is a checker for SV-COMP.

The obvious question at this point is why the checkers should be trusted. In other words, what kind of additional guarantee do we get
by using these checkers instead of just relying upon the deduction tools? 
On the positive side, these checkers are usually less complex programs compared with full deduction tools, since they only have to check existing
proofs instead of generating proofs, and therefore are less likely to contain bugs. Moreover, since often the checkers are independent of
the deduction tools, using such a checker can increase the level of trust, since then two independent systems agree upon the same proof.

\subsection{Certification with Proof Assistants via Proof Script Generation}
\label{subsect:certification-with-proof-assistants}

To increase the level of trust even further, we want to profit from the high reliability of proof assistants
with their small trusted code base (Section~1.6). 
Moreover, using proof assistants not only we can increase the reliability with respect to\ question~\ref{appl}, but
we can also improve it with respect to\ question~\ref{thms}, namely, by formally verifying the theorems. 
  
We illustrate a workflow for validating both questions with the help of a proof assistant by applying it on \exref{ex ackermann}. 
The idea is to develop a library, generate a proof script for each example certificate, and then run the proof assistants to check
the scripts, as shown in Figure~\ref{fig:certification}.

\newcommand\opx{0}
\newcommand\opw{1.2cm}
\newcommand\fix{2}
\newcommand\cox{3.2}
\newcommand\cow{7.2cm}
\begin{figure}[t]
\begin{tikzpicture}
[operation/.style={rounded corners=.1cm,rectangle,fill=black!10,thick},
 file/.style={rectangle,draw=black,thick},
 content/.style={anchor=west,rectangle,draw=black}]
\node[operation] (op) at (\opx,1) {Operation};
\node[file] (file) at (\fix,1) {File};
\node[content] (Content) at (\cox,1) {Example Content};
\node[operation] (fld) at (\opx,0) {1. Formalization};
\node[file] (fl) at (\fix,0) {Library};
\node[content] (fc) at (\cox,0) {\parbox{\cow}{\texttt{definition rewrite-relation R = ...\\
  theorem \ref{LPO sound}: SN > --> red-order (LPO >)}}};
\draw[->] (fld) -> (fl);
\draw[thick] (-1.05,0.6) -- ++(\textwidth,0);

\node[file] (if) at (\fix,-1) {Input};
\node[content] (ic) at (\cox,-1) {\parbox{\cow}{\texttt{ack(0,m) -> m; ...}}};

\node[operation] (ded) at (\opx,-1.5) {\parbox{\opw}{Deduction\\Tool}};

\node[file] (pf) at (\fix,-2) {Property};
\node[content] (pc) at (\cox,-2) {\parbox{\cow}{\texttt{Terminating}}};
\node[file] (cf) at (\fix,-2.5) {Certificate};
\node[content] (cc) at (\cox,-2.5) {\parbox{\cow}{\texttt{LPO(precedence: ack > s)}}};

\draw[->] (if) to [out=180,in=90] (ded);
\draw[->] (ded) to [out=270,in=180] (pf);
\draw[->] (ded) to [out=270,in=180] (cf);

\node[operation] (sgen) at (\opx,-3.3) {\parbox{\opw}{2. Script\\Generator}};
\node[file,anchor = north] (sf) at (\fix,-3.5) {\parbox{1.4cm}{Script with\\
  property + \\
  proof}};
\node[content,anchor = north west] (sc) at (\cox,-3.5) {\parbox{\cow}{\texttt{%
definition ack-trs = ... \\
lemma SN (rewrite-relation ack-trs) \\
apply (theorem \ref{redorders} where $\succ$ = LPO)\\
...
}}};

\coordinate (ingen) at (2.8,-2.8);

\draw (if) to [out=0,in=90] (ingen);
\draw (pf) to [out=0,in=90] (ingen);
\draw (cf) to [out=0,in=90] (ingen);
\draw[->] (ingen) to [out=270,in=90] (sgen);

\draw[->] (sgen) to [out=270,in=180] (sf);

\node[operation] (check) at (\opx,-5.3) {\parbox{\opw}{3. Proof\\Checking}};

\coordinate (incheck) at (3,-5);
\coordinate (preincheck) at (3,-1);
\coordinate (preincheck2) at (3,-4.5);
\draw (sf.east) to [out=0,in=90] (preincheck2);
\draw (fl) to [out=0,in=90] (preincheck);
\draw (preincheck) to (incheck);
\draw[->] (incheck) to [out=270,in=0] (check);

\end{tikzpicture}
\caption{Certification Workflow with Proof Assistants and proof script Generation
\label{fig:certification}}
\end{figure}
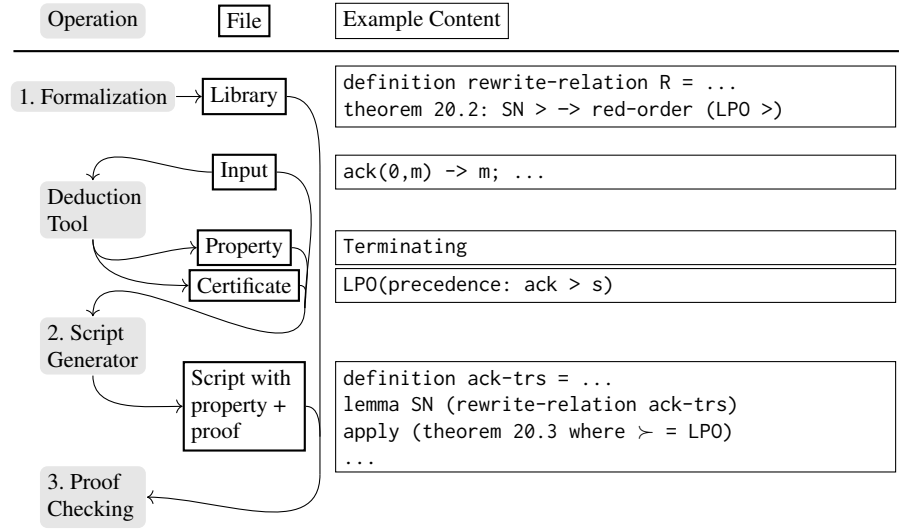

Let us illustrate all three steps in some more detail.

\begin{enumerate}
\item\label{lib} We need to develop a library that formalizes several domain-specific notions and theorems.
  In the LPO\index{LPO} example, this library will contain notions such as strong normalization and term rewriting,
  the LPO might be formally defined as inductive predicate, and LPO properties
  such as \thmref{LPO sound} are available
  as formal theorems. However, the library does not include any particular inputs and
  certificates; for example, it contains neither the rewrite system nor the precedence of
  \exref{ex ackermann}.

\smallskip

\item\label{gen}
  We need to develop a proof script generator:\ its task is to transform a
  given input certificate into a proof script for the proof assistant.
  This script contains two essential parts:\ the definition of the property of interest
  and the formal proof of this property. For \exref{ex ackermann},
  the first part will
  contain the formal definition of the Ackermann--Péter system and the statement that it
  is terminating.
  The second part will be a formal proof of this statement,
  obtained by encoding the proof of the certificate
  into a script. This script may rely on everything that is provided by the proof assistant, the notions and theorems from the formal library of Step~\ref{lib}, as well as on tactics that
  have been tailor-made for the certification task.

\smallskip

\item\label{check} Finally, checking the proof is possible:\
  a user must validate that the correct property has been defined in the first part,
  and then verifying the proof of the property is achieved by running the proof assistant.
\end{enumerate}

For the formal definition of the property, there are several design decisions required. For example,
for the Ackermann--Péter system one could first define a new datatype for the given signature (with constructors
$\Fack,\Fs,\Fz$) or just encode the three symbols by strings.
The first approach has the advantage that the proof assistant's type checker
can ensure that no garbage terms
such as $\Fack(\Fz,\Fnil)$ can occur, at the cost of having to create a symbol type for
every input. Similarly, one can use invariants or dependent types to ensure that all terms respect the arities of
the symbols, forbidding an input term such as $\Fack(\Fs)$.


After having defined the property, we need a way to certify $\ell \succ r$ for all rules of the rewrite system,
where we mention three alternatives.
First, when given the extended certificate described in \exref{ex ackermann},
the proof script generator can easily transform the inference tree into a proof script
where each inference rule is translated into the corresponding
introduction rule of the formal version of LPO. Second, when only the basic certificate is available,
the generator might reconstruct the inference tree for each rewrite rule with the help
of some unverified algorithm,
and then translate the tree as in the first case; of course, adding extra information
into the certificate can also be done by an unverified preprocessor
that runs before the proof script generator. Alternatively, the proof script generator may just invoke
some tailor made tactic, that internally reconstructs the inference trees and additionally applies the corresponding introduction
rules.
Third, one can perform a deep embedding\footnote{Modeling properties and proofs can
be done using either a deep embedding or a shallow embedding. We do not explain these concepts
here, but just refer to Section~5.6.1 
for further details.} and implement a verified algorithm
within the proof assistant, which either ensures $\ell \succ r$ for a given precedence
or outputs some error message.
Then the proof script generator only needs to invoke that algorithm and check
that the result is not an error.

All of these three options can also be used to check complex combined proofs, e.g.,
a termination proof consisting of several different termination techniques.

Overall, certification with proof script generation has the advantage that it
significantly decreases the trusted code base
and in particular covers both questions~\ref{thms} and \ref{appl}.
One disadvantage in comparison with the usage of trusted checkers as in Section~\ref{subsect:trusted-checkers} 
is a reduced 
execution speed: proof script generation in combination with running the proof assistant 
is usually much slower than just executing the trusted checker.
This might be a significant problem when it comes to checking proofs which require 
many computations.

\subsection{Certification with Proof Assistants via External Certification}
\label{subsect:external-certification}

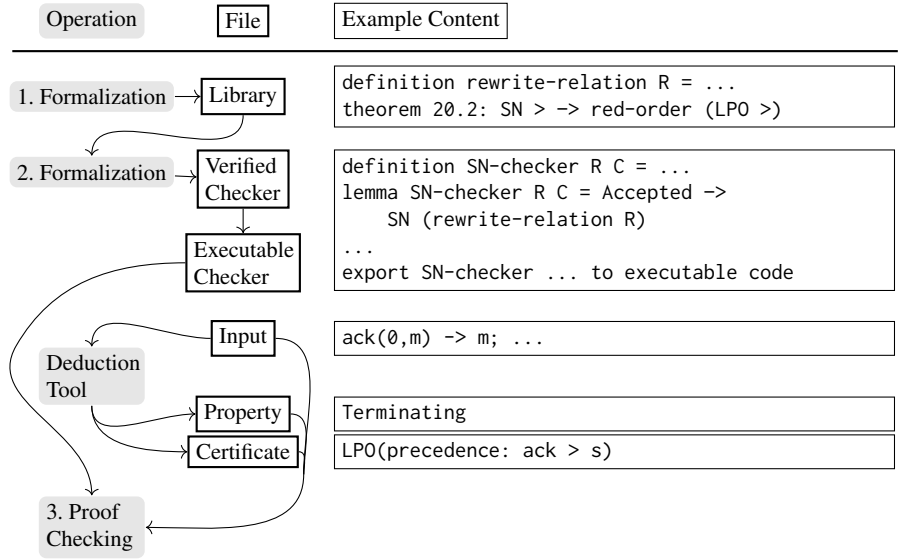
\begin{figure}[t]
\begin{tikzpicture}
[operation/.style={rounded corners=.1cm,rectangle,fill=black!10,thick},
 file/.style={rectangle,draw=black,thick},
 content/.style={anchor=west,rectangle,draw=black}]
\node[operation] (op) at (\opx,1) {Operation};
\node[file] (file) at (\fix,1) {File};
\node[content] (Content) at (\cox,1) {Example Content};
\node[operation] (fld) at (\opx,0) {1. Formalization};
\node[file] (fl) at (\fix,0) {Library};
\node[content] (fc) at (\cox,0) {\parbox{\cow}{\texttt{definition rewrite-relation R = ...\\
  theorem \ref{LPO sound}: SN > --> red-order (LPO >)}}};
\draw[->] (fld) -> (fl);
\draw[thick] (-1.05,0.6) -- ++(\textwidth,0);

\node[operation] (co) at (\opx,-1) {2. Formalization};
\node[file,anchor = north] (cf) at (\fix,-0.7) {\parbox{1cm}{Verified\\
  Checker}};
\node[content,anchor = north west] (cc) at (\cox,-0.7) {\parbox{\cow}{\texttt{%
definition SN-checker R C = ... \\
lemma SN-checker R C = Accepted -->\\
\phantom{XXXX}SN (rewrite-relation R)\\
...\\
export SN-checker ... to executable code
}}};

\draw[->] (fl) to[out=270, in = 90] (co);
\draw[->] (co) to (cf);

\node[file] (ecf) at (\fix,-2.2) {\parbox{1.3cm}{Executable\\
  Checker}};

\draw[->] (cf) to (ecf);

\node[file] (if) at (\fix,-3.2) {Input};
\node[content] (ic) at (\cox,-3.2) {\parbox{\cow}{\texttt{ack(0,m) -> m; ...}}};

\node[operation] (ded) at (\opx,-3.7) {\parbox{\opw}{Deduction\\Tool}};

\node[file] (pf) at (\fix,-4.2) {Property};
\node[content] (pc) at (\cox,-4.2) {\parbox{\cow}{\texttt{Terminating}}};
\node[file] (cf) at (\fix,-4.7) {Certificate};
\node[content] (cc) at (\cox,-4.7) {\parbox{\cow}{\texttt{LPO(precedence: ack > s)}}};

\draw[->] (if) to [out=180,in=90] (ded);
\draw[->] (ded) to [out=270,in=180] (pf);
\draw[->] (ded) to [out=270,in=180] (cf);

\node[operation] (check) at (\opx,-5.7) {\parbox{\opw}{3. Proof\\Checking}};

\coordinate (incheck1) at (-1,-3.3);
\coordinate (incheck2) at (2.8,-5);

\draw (ecf) to [out=180,in=60] (incheck1);
\draw[->] (incheck1) to [out=240,in=90] (check);

\draw (if) to [out=0,in=90] (incheck2);
\draw (pf) to [out=0,in=90] (incheck2);
\draw (cf) to [out=0,in=90] (incheck2);
\draw[->] (incheck2) to [out=270,in=0] (check);

\end{tikzpicture}
\caption{External certification workflow}
\label{fig:external certification}
\end{figure}

A solution to the problem of reduced execution speed of the proof script approach is available
if the proof scripts are always of a very specific shape, namely, they
just consist of the invocation of some verified algorithm.
In this case, the algorithm provides a sufficient criterion for
the desired property. Here, the certificate is mainly used to guide the verified algorithm, e.g., which techniques
should be applied, and how are these parameterized.
Given such a verified algorithm, certification is possible outside the proof assistant,
which gives an alternative workflow to Steps~\ref{lib}--\ref{check} mentioned in Section~\ref{subsect:certification-with-proof-assistants}:\ \emph{external certification},
shown in Figure~\ref{fig:external certification}.

\begin{enumerate}[label=\arabic*$'$., ref=\arabic*$'$]
\item\label{lib'} Same as \ref{lib}.

\smallskip

\item\label{algs} For each supported property $P$ and certificate $C$, provide a verified algorithm
  that is a sufficient condition for the property:\ whenever the algorithm on input~$I$ and $C$
  does not complain, then $P\ I$ holds. Export the verified algorithms into some executable program.

\smallskip

\item\label{run} Run the program on inputs and certificates. Here some wrapper can be used to select
  the suitable verified algorithm depending on the property $P$ that needs to be verified.
\end{enumerate}

Clearly, external certification is faster than the previous approach:\ running a program
is much faster than first starting a proof assistant and then letting it check a proof script.
For instance, in experiments about termination proof certification 
the certifier based on external certification was 50 times faster than its competitors that have
been using proof scripts \cite{DBLP:conf/tphol/ThiemannS09}.
As a consequence, more certificates can be handled in reasonable time. This is in particular the case if the
certification task requires many \emph{computations}\index{computation}.

Note that it is not necessarily advantageous or always possible
to provide helpful details about computations in the certificate.
For example, for checking safety proofs via tree automata (a $\mathbf{PSPACE}$-hard problem),
one has to check a certain property
for \emph{all} state-substitutions. Here, listing these state-substitutions in the certificate
is not advantageous at all:\ one has to generate all state-substitutions in a verified way in order to
satisfy the criterion that ensures safety; consequently, comparing the generated substitutions with 
the ones in the certificate just poses some
extra work; finally, certificates with an explicit list of substitutions would be much more bulky and 
are thus more tedious to generate.

Hence, for some kinds of proofs, efficiency of the verified algorithms becomes a critical factor.
In these cases, the integration of optimizations into the verified algorithms becomes essential. 
This integration is often done
using a refinement\index{refinement} approach; that is, first an abstract algorithm is proved correct, and then
it is refined to an executable one where efficient data structures and algorithms are used
to implement the abstract operations.

Besides improved efficiency, external certification has one further advantage in comparison with
proof script generation: it is less brittle in the face of changes in the proof assistant. The
proof assistant only needs to be integrated statically for Steps~\ref{lib'} and
\ref{algs}. By contrast, such changes might also require modifications
of the proof script generator in Step~\ref{gen} and required adaptations only become visible
when running it on example certificates.

However, external certification also has two disadvantages compared with proof script generation.
First, some design decisions, which have been discussed before, are no longer available;
for example, we cannot create a datatype for the signature that is used in a certificate,
and also a shallow embedding is not possible anymore, since the verified algorithm runs outside
the proof assistant. 
The second disadvantage is that external certification has a larger trusted code base.
For instance, external certification invokes the code generator\index{code
generation} of a proof assistant and an external compiler, and both of these
tools might not be formally verified. We refer to Section~5.6 
for further details on code generation.

\subsection*{Summary on Certification}

To conclude, we mention several successful examples of certification using proof assistants
as it was described in Sections~\ref{subsect:certification-with-proof-assistants}
and \ref{subsect:external-certification}.
\begin{itemize}
\item Coccinelle\index{Coccinelle} is a library on rewriting for the Rocq Prover (formerly known as Coq),
  and CiME3\index{CiME3} contains a corresponding proof script generator
  \cite{DBLP:conf/frocos/ContejeanCFPU07,DBLP:conf/rta/ContejeanCFPU11}.
  Their approach uses a
  mixture of shallow and deep embedding and supports several termination techniques for term rewriting.

\smallskip

\item CoLoR\index{CoLoR} is another Rocq libary on rewriting, and initially Rainbow\index{Rainbow}
  was used as the corresponding proof script
  generator \cite{DBLP:journals/mscs/BlanquiK11,DBLP:journals/aaecc/Koprowski09}. In contrast to Coccinelle\slash CiME3,
  most of the formalization is using a deep embedding, so that later on external certification of termination
  proofs became available, and the development of Rainbow was stopped.

\smallskip

\item IsaFoR\index{IsaFoR} is the Isabelle Formalization of Rewriting \cite{DBLP:conf/tphol/ThiemannS09}.
  From the outset, it was designed for external certification. Besides termination proofs of rewrite
  systems, it also supports confluence proofs, complexity proofs, completion proofs, and proofs
  about tree automata and integer transition systems.

\smallskip

\item The developers of the three certifiers above have agreed on a
  standard certification format, CPF\index{CPF} \cite{DBLP:journals/corr/SternagelT14}. This format
  is currently used in the annual termination competition (termCOMP) and confluence competition
  (CoCo) to validate results of competing certificate-producing tools.

\smallskip

\item Certifiers for SAT\index{SAT solving} have been
  verified in various proof assistants \cite{DBLP:conf/tacas/TanHM21,DBLP:conf/cade/Cruz-FilipeHHKS17,DBLP:journals/jar/Lammich20}.
  Here, the interesting part is of course checking unsatisfiability proofs.
  All cited certifiers expect their input in the DRAT\index{DRAT} format \cite{DBLP:journals/corr/Heule16},
  a format that is easy to produce for tools and that has been
  used in recent SAT competitions.
  Since efficiency is crucial for checking unsatisfiability proofs,
  all approaches are based on external certification.
  Moreover, preprocessing of the DRAT certificates is done
  by unverified algorithms, e.g., by adding extra information to speed up
  unit propagation in the verified algorithms.

\smallskip

\item Timbuk\index{Timbuk} is a collection of tools that use tree automata completion
  to verify higher-order functional programs or cryptographic protocols.
  To increase reliability of the unverified Timbuk, a verified certifier
  for tree automata completion has been developed in Rocq \cite{DBLP:conf/cade/BoyerGJ08}.
  Since checking the properties of tree automata involve heavy computations,
  the verified checker uses external certification.
\end{itemize}

\index{certification|)}

\section{Extensions of the Inference Systems}

When running certifiers on certificates of deduction tools, it might happen
that some certificates are rejected. This in particular happens when a fixed combination of
technique, certifier, and deduction tool is tested for the first time, e.g., directly after a new technique
has been added to the certificate format, or when some tool added certificate generation for the first time.

There are at least four different reasons for rejected certificates.

First, there might be some misunderstanding or imprecision in the certificate format.
For example, should the indices of subterms $s_i$ of inference rule (\ref{sub})
be started from 1 (as usual in papers) or from 0 (as usual in implementations)?
Such problems will result in many rejected certificates and 
are therefore easy to detect. Moreover, they can be repaired easily, e.g., 
by making the specification of the certification format more precise. Afterward, one
just has to adjust 
either the parser of the certifier or the certificate generation part of the deduction tool.
Here, a close collaboration between the developers of the certifier and the deduction tool
is definitely helpful, where ideally the certificate format is codeveloped right from the start.

Second, there might be a bug in the parser of the certifier or in the certificate generating part
of the deduction tool. Problems of this kind are very similar to the first one: they are usually
detected immediately and are easy to repair.

Third, a certificate might contain an invalid proof, so certification reveals a real bug
in the deduction tool. In this situation, the bug
can be repaired in the tool, possibly turning an unsound tool into a sound one.
We briefly mention three examples of these bugs. All of them remained unnoticed for some years and got
revealed as soon as certification of the effected techniques became available.
The first bug caused wrong nontermination answers, i.e., the tool ``disproved termination'' for some
terminating rewrite systems. The simple reason was that some of the applied techniques in the tool ignored the evaluation strategy
of the rewrite systems \cite{DBLP:conf/itp/SternagelT12}.
The second bug resulted in wrong termination proofs \cite{DBLP:conf/itp/Thiemann13}. Here the reason was a mistake
in the implementation of some inference system for manipulating first-order formulas.
Finally, we mention bugs in tools for complexity analysis that have been detected with the help of certification \cite{DBLP:conf/rta/AvanziniST15}.

Fourth, a certificate might contain proof steps that, despite being valid in
principle, are rejected by the certifier. The reason might be that the certifier
uses only a sufficient criterion to ensure validity of a certain proof step, or that the
deduction tool uses an extended set of inference rules in its implementation
that deviates from what is described in a corresponding paper, and hence also
from what is checked with respect to the formal statements.
In both cases, an easy solution is to integrate an option into
the deduction tool so that users can restrict the reasoning engine of the tool
in a way that it solely uses approximations and inference rules that are supported
by the certifier.
Such a restriction will usually come with a decrease in power. The alternative
is to make the certifier more powerful, e.g., by adding the new inference
rules to the formal definition.

As an example, consider an extension of LPO by a new inference rule that supports
least elements where it is allowed to compare a variable $x$ against a constant $c$
that has least precedence.
To this end, we can define a new nonstrict
LPO order $\succsim$ satisfying ${\succ} \subseteq {\succsim}$ and including rule (\ref{least}), and modify the
existing rule (\ref{lex}). In this way, rules such as $\Ff(x,\Fs(y),z) \to \Ff(\Fz,y,x)$ can
be proved terminating by LPO.
\begin{gather*}
\label{least}\iirule{\Forall{d}{d \geq c}}{x \succsim c}{} \tag{least} \\
\iirule{\Forall{j \in \{1,\dots,i-1\}}{s_j \succsim t_j \qquad s_i \succ t_i} \qquad
  \Forall{j \in \{i+1,\dots,n\}}{f(s_1,\dots,s_n) \succ t_j}}%
{f(s_1,\dots,s_n) \succ f(t_1,\dots,t_n)}{} \tag{lex$'$}
\end{gather*}

The advantage of having a formal proof at this point is that one can easily spot where
the proofs of the properties of LPO need adjustments.
By contrast, doing a rigorous proof checking manually on paper would
at least be tedious, so it might just be sketched, and errors might remain hidden.
For example, while the above extension of LPO is sound, the very same extension of
RPO is unsound, since then a crucial property of a reduction order is no longer satisfied
\cite{DBLP:conf/rta/ThiemannAN12}.

Knuth--Bendix orderings\index{Knuth--Bendix ordering} compatible with associativity and commutativity,
called AC-KBO, provide
another example that an extension or a modification of a set of inference rules is error-prone
when done manually.
There are several variants of AC-KBO available
\cite{DBLP:conf/alp/Steinbach90,DBLP:conf/cade/KorovinV03,DBLP:journals/tplp/0002WHM16},
for which contradictory properties are stated
in various papers:\
for example, one paper states that a particular variant is flawed, and a later paper
claims that there is no flaw in that variant, but instead one only needs an alternative proof.
More details on the conflicting results on AC-KBO are given by Yamada et al.~\cite{DBLP:journals/tplp/0002WHM16}, and at least their
variant of AC-KBO has been verified in Isabelle\slash HOL~\cite{DBLP:conf/cpp/LochmannS19}, so that
the issue is finally settled via formal verification.

Another success story is part of IsaFoL\index{IsaFoL}, the Isabelle Formalization of Logic \cite{DBLP:conf/cpp/Blanchette19},
that aims at developing formal theories about logics, proof systems, and automatic provers.
It contains a formal soundness and completeness proof~\cite{DBLP:conf/cade/SchlichtkrullBT18}
of a first-order prover as it is presented in
the \emph{Handbook of Automated Reasoning}~\cite{DBLP:books/el/RV01/BachmairG01}.
Besides the intrinsic merits of verifying a prover,
it is now possible to safely explore or modify
the underlying inference
rules of the calculus (``hack without fear'' as an anonymous reviewer wrote \cite{DBLP:conf/cpp/Blanchette19}),
since the proof assistant will rigorously check all proofs again and point to those properties which
break or need adjustments in their proofs.

\section{Verified Deduction Tools}
\label{sec:verified-deduction-tools}

We saw in Section~\ref{sect:certification}
that certification can be used to validate the output of deduction tools.
Such a validation is no longer required if the deduction tool itself
has been fully verified. However, it is more work to
develop a verified tool than a verified certifier.
One of the reasons is that deduction tools are more efficiency critical 
than checkers, since the former tools have to search for a proof. 
Consequently, they usually contain more optimizations, which then all need to be formally verified.
To this end, often one can reuse commonly used algorithms and data structures (e.g., tree-based dictionaries, sorting
algorithms) since verified implementations are already contained in libraries of proof assistants.
A problem occurs if  the deduction tool to be verified
crucially relies upon an imperative algorithm with mutable states. In that case the verification and modeling task might
become more complicated, in particular if the default modeling language of the proof assistant is purely functional.

A successful example for a verified deduction tool is the IsaSAT\index{IsaSAT} solver of Fleury et al.~\cite{DBLP:journals/jar/BlanchetteFLW18}
that has been developed
as part of IsaFoL\index{IsaFoL}. It starts with a high-level description of the algorithm in the form of 
inference rules which are both abstract and non-deterministic. At this point many design decisions are still open and 
implementation-dependent optimizations are not yet included, in particular since these are not expressible on the abstract level. 
Afterward, an executable and deterministic verified implementation is developed from the inference rules by a series of refinement\index{refinement} steps. 
One
of the refinements is the inclusion of two watched literals
to speed up unit propagation and conflict detection.
Interestingly, during the formal proof development it was detected that
the stated (paper) invariant for two watched literals needed adjustments.
A fixed invariant is now available in the formal proof as well as
in its textual description \cite[Section~6]{DBLP:journals/jar/BlanchetteFLW18}.

It should be said that at the time of writing,
the combination of a highly optimized deduction tool with a certifier run afterward is usually more efficient than
a fully verified deduction tool, despite the overhead in the certified approach where the
proof is processed twice.
This was the approach taken by Lammich for his highly optimized unsatisfiability
certificate checker~\cite{DBLP:journals/jar/Lammich20}.
However, the fully verified tool might have stronger formal guarantees (e.g., being a decision procedure),
whereas in the certified approach the deduction tool may contain bugs and therefore might produce uncertifiable proofs.

Last but not least, proof assistants themselves
are also deduction tools and are therefore subject to verification.
The work on the verification of Rocq and its underlying metalogic\index{metalogic} have
already been mentioned in Section~18.3.2.
As further examples, we briefly
mention work on verifying HOL
Light\index{HOL Light}~\cite{DBLP:conf/cade/Harrison06,DBLP:conf/itp/AbrahamssonMKS22} and work on
Milawa\index{Milawa}~\cite{DBLP:journals/jar/DavisM15}, a verified proof assistant
that is modeled after ACL2\index{ACL2}. Furthermore, also certification is possible: 
proof assistants itself can export their internally generated proofs as a certificate,
which can then be independently checked, e.g., by Dedukti\index{Dedukti} \cite{DBLP:conf/fscd/HondetB20}.

\section{Conclusion}

We have seen how deduction tools can profit from the usage of proof assistants. 
Verified certifiers detected several bugs in implementations of deduction tools,
where some of these bugs have not been noticed for years. Moreover, certifiers are 
successfully applied in competitions of deduction tools, where they are used to 
validate a large amount of automatically generated proofs.

Core techniques that are implemented in deduction
tools have been formally verified, resulting in two further benefits. 
On the one hand, faulty proofs in published papers
could be revealed and repaired. On the other hand, extensions of these 
techniques can safely be investigated with the help of a proof assistant: 
``hack without fear.''

Finally, we also mentioned the development of verified deduction tools. 
Here, the cost of verification is usually much higher than for certifiers,
but it can be done. Although the unverified tools usually outperform
the verified tools, recently it was shown that the situation might change:
the verified solver IsaSAT\index{IsaSAT} won in the EDA-Challenge 2021 against its
unverified competitors. It will be interesting to see future developments
of verified solvers, in particular also those that target
more complex properties than SAT.

{
\raggedright

}

\let\comment=\oldcomment

\end{document}